\documentclass[12pt,a4paper]{article}

\def\fun#1#2{\lower3.6pt\vbox{\baselineskip0pt\lineskip.9pt
\ialign{$\mathsurround=0pt#1\hfil##\hfil$\crcr#2\crcr\sim\crcr}}}

\title{$B \to \pi \pi$ decays: branching ratios and CP asymmetries}
\author{A.B. Kaidalov\thanks{kaidalov@itep.ru} \  and
M.I. Vysotsky\thanks{vysotsky@itep.ru} \\ Institute of Theoretical
and Experimental Physics, \\ Moscow, Russia}
\date{}

\begin{document}

\maketitle

\abstract{Theoretically motivated smallness of the penguin
amplitude in $B \to \pi \pi$ decays allows to calculate the value
of the unitarity triangle angle $\alpha (\phi_2)$ with good
accuracy. The relatively large branching ratio of the decay into
$\pi^0 \pi^0$ is explained by the large value of FSI phase
difference between  decay amplitudes with $ I = 0$ and $ I = 2$ .}

\bigskip

PACS: 12.15.Hh, 13.20.He

\large

\section{Introduction}

The exclusive decay amplitudes of hadrons are determined by dynamics
at large distances and can not be calculated with presently
available  QCD tools.

Fortunately it was found long ago that the experimental data on branching
ratios and CP asymmetries of $B \to \pi \pi$ decays allow to
determine the value of the unitarity triangle angle $\alpha$ with
essentially no hadronic  input using isospin invariance of
strong interactions only \cite{GL}. However, large
experimental uncertainties in particular in the values of the direct
CP asymmetries lead to poor accuracy in the value of $\alpha$
determined in this way.

If the penguin amplitudes are negligible in charmless strangeless
$B$ decays we would determine the value of unitarity triangle
angle $\alpha$ from CP asymmetry $S_{+-}$ extracted from
 $B \to \pi^+ \pi^-$ decay data with essentially no
theoretical uncertainties.
As it was found in paper \cite{OV} neglecting penguin
amplitudes one gets the values of
angle $\alpha$ from CP asymmetries in $B_d$ decays to $\pi^+ \pi^-$,
 $\rho^+ \rho^-$ and  $\pi^{\pm} \rho^{\mp}$
 consistent with the global fit of
unitarity triangle.  Since the penguin contributions to these
decays are different \cite{Buch} the fact that the numerical
values of $\alpha$ are close to each other testifies in favor of
smallness of penguin amplitudes. Small penguin corrections to
these decay amplitudes were accounted for in \cite{V} where the
hadronic amplitudes were found from the quark amplitudes with the
help of factorization. However, it is well known that the
branching ratio of $B_d (\bar B_d) \to \pi^0 \pi^0$ decay
predicted by factorization appears to be more than 10 times
smaller than the experimental data. The way out of this
contradiction could be large FSI phases in  $B \to \pi \pi$
decays. The validity of this theoretical ingredient will be
checked by the more accurate experimental data.

Though the penguin contribution is relatively small compared to
tree amplitudes and can be neglected in the first approximation in
the decay probabilities and in the CPV parameters $S$ it
determines the CPV parameters $C$ and should be accounted for in
the analysis of the complete set of observables.

The charmless strangeless B decays  are described by $b \to u \bar
u d$ quark transition. The effective Hamiltonian responsible for
this transition consists of two parts: the tree level weak
amplitude (operators $O_1$ and $O_2$ in standard notations)
dressed by gluons  and the gluon penguin amplitudes (operators
$O_3 - O_6$); the parametrically small electroweak penguins are
omitted. The gluon penguins being very important in $\Delta S = 1$
strange particles nonleptonic weak decays are almost negligible in
$\Delta B = 1$, $\Delta S = 0$ transitions. The reason is twofold:
firstly, Wilson coefficients are much smaller in case of B decays
because infrared cutoff is at $\mu \sim m_b$ instead of $\mu \sim
\Lambda_{QCD}$; secondly, the enhancement factor originated from
the right-handed currents $m_\pi^2/m_s(m_u + m_d) \sim 10$ for
strange particles decays is replaced by $m_\pi^2/m_b(m_u + m_d)
\sim 1/3$ for beauty hadrons. That is why after presenting the
general phenomenological expressions for the amplitudes we will
start our analysis of $B \to\pi\pi$ decays in Section 2 by the
sequestered  Hamiltonian which does not contain penguin
contributions \footnote{ Let us stress that while from the
smallness of $B \to \pi^0 \pi^0 $ decay width it would follow that
penguins are small, the opposite statement is not correct: the
relatively large width to neutral pions does not necessary mean
that penguins are large.}. From the experimental data on $B_d
(\bar B_d) \to\pi^+\pi^-, \pi^0 \pi^0$ and $B_u \to\pi^+ \pi^0$
branching ratios we will extract the moduli of the amplitudes of
the decays into $\pi\pi$ states with isospin zero $A_0$ and two
$A_2$ and find the final state interaction (FSI) phase shift
$\delta \equiv \delta_2 - \delta_0$ between these two amplitudes.
The value of the unitarity triangle angle $\alpha$ in this
approximation is directly determined by  CP asymmetry $S_{+-}$.

While the absolute values of the amplitudes $A_0$ and $A_2$ are
reproduced with good accuracy by the factorization formulas, the
FSI phase shift appears to be unexpectedly large, $\delta = -(53^o
\pm 7^o)$. This is the reason why $B \to \pi^0 \pi^0$ decay
probability is significantly enhanced in comparison with the naive
factorization approach, where one neglects $\delta$. In Section 3
FSI phase differences in $K \to \pi\pi$, $D \to \pi\pi$ and $B \to
D\pi$ decays are considered. In all these cases the phases are
large, which is attributed to the existence of $I=0$ resonances in
$\pi\pi$ scattering in the cases of $K \to\pi\pi$ and (partly) in
$D\to\pi\pi$ decays while large FSI phases in $B\to D\pi$ decays
are unexpected. In Section 4 we consider the theoretical estimates
of $\delta$ and show how FSI can enhance B width to neutral pions
not enhancing that to neutral $\rho$ mesons in accordance with
experimentally observed suppression of $B \to \rho^0 \rho^0$ decay
width.

In Section 5 the penguin contributions are considered; the
corrections to the numerical values of $A_0$ and $\delta$ due to
gluon penguin amplitudes are determined, as well as the correction
to the unitarity triangle angle $\alpha$ and the values of CP
asymmetries $C_{+-}$ and $C_{00}$. In Conclusions the pattern of
the $B\to\pi\pi$ decay amplitudes emerging from the experimental
data is presented. Appendix contains the calculations of the decay
amplitudes in factorization approximation.

\section{$B\to\pi\pi$ without penguins: decay amplitudes from
branching ratios}

The quark Hamiltonian responsible for $B\to\pi\pi$ decays has the parts
with $\Delta I = 1/2$ and $\Delta I = 3/2$ which produce
$\pi$-mesons in the states with $I=0$ and $I=2$ correspondingly.
QCD penguins
having $\Delta I = 1/2$ contribute only to the $I = 0$ amplitude.
Taking into account the corresponding Clebsch--Gordan coefficients and
separating the penguin contribution ($P$) with the
CKM phase different from that of $A_0$
 we obtain:
\newpage
\begin{eqnarray}
M_{\bar B_d \to \pi^+\pi^-} & = & \frac{G_F}{\sqrt 2}|V_{ub}
V_{ud}^*| m_B^2 f_\pi f_+(0) \left\{e^{-i\gamma}\frac{1}{2\sqrt
3}A_2 e^{i\delta_2} \right. + \nonumber \\ & + & e^{-i\gamma}
\left.\frac{1}{\sqrt 6} A_0 e^{i\delta_0}
+ \left|\frac{V_{td}^*
V_{tb}}{V_{ub} V_{ud}^*}\right| e^{i\beta} P
e^{i(\delta_p + \tilde\delta_0)}\right\} \;\; , \label{1}
\end{eqnarray}
\begin{eqnarray}
M_{\bar B_d \to \pi^0\pi^0} & = & \frac{G_F}{\sqrt 2}|V_{ub}
V_{ud}^*| m_B^2 f_\pi f_+(0) \left\{e^{-i\gamma}\frac{1}{\sqrt
3}A_2 e^{i\delta_2} \right. - \nonumber \\ & - & e^{-i\gamma}
\left.\frac{1}{\sqrt 6} A_0 e^{i\delta_0} - \left|\frac{V_{td}^*
V_{tb}}{V_{ub} V_{ud}^*}\right| e^{i\beta} P
e^{i(\delta_p  + \tilde\delta_0)}\right\} \;\; , \label{2}
\end{eqnarray}
\begin{equation}
M_{\bar B_u \to \pi^-\pi^0} = \frac{G_F}{\sqrt 2}|V_{ub} V_{ud}^*|
m_B^2 f_\pi f_+(0) \left\{\frac{\sqrt 3}{2\sqrt 2} e^{-i\gamma}
A_2 e^{i\delta_2} \right\} \;\; , \label{3}
\end{equation}
where $V_{ik}$ are CKM matrix elements and the penguin amplitude
with an intermediate $c$-quark multiplied by $V_{ub} V_{ud}^* +
V_{cb} V_{cd}^* + V_{tb} V_{td}^* = 0$ is subtracted from the
penguin amplitudes with intermediate $u$-, $c$- and $t$-quarks
(the so-called t-convention) \footnote{We prefer t-convention
since the penguin contribution to the amplitude $A_0$ in it is
suppressed as $m_c^2/(m_b^2 \ln m_W^2/m_b^2)$, while in the
$c$-convention one subtracts the penguin amplitude with
intermediate $t$-quark thus making the penguin contribution to
$A_0$ comparable to the penguin term.}. To check if the
factorization works in $B \to \pi \pi$ decays it is convenient to
introduce $f_+(0)$ - the value of the formfactor which enters the
amplitude of semileptonic $B_d \to \pi l\nu$ decay at zero
momentum transfer in Eqs. (1)-(3). $\gamma$ and $\beta$ are the
angles of the unitarity triangle; $\delta_2$ and  $\delta_0$ are
 FSI phases of the tree
amplitudes with $I=2$ and $I=0$
(below we will use $\delta \equiv \delta_2 - \delta_0$),
 $\delta_p$ originates from
the imaginary part of the penguin loop with $c$-quark propagating
in it \cite{BSS} while $\tilde\delta_0$ is long distance FSI phase
of the penguin amplitude.
$\tilde\delta_0$ in general is different from $\delta_0$; in Section 4
we will argue that $\rho \rho $ intermediate state generate large
value of $\delta_0$ while its contribution into
$\tilde\delta_0$ is smaller: (pseudo)scalar part of penguin operator
do not produce $\rho$ mesons.

The charge conjugate amplitudes are obtained by the same formulas
with substitution $\beta, \gamma \to -\beta, -\gamma$.

The CP asymmetries are given by \cite{G}: $$ C_{\pi\pi} \equiv
\frac{1-|\lambda_{\pi\pi}|^2}{1+|\lambda_{\pi\pi}|^2} \; , \;\;
S_{\pi\pi} \equiv \frac{2{\rm Im}(\lambda_{\pi\pi})}{1+
|\lambda_{\pi\pi}|^2} \; , \;\; \lambda_{\pi\pi} \equiv
e^{-2i\beta} \frac{M_{\bar B \to \pi\pi}}{M_{B\to\pi\pi}} \;\; ,
$$ where $\pi\pi$ is $\pi^+\pi^-$ or $\pi^0 \pi^0$.

Now we have all the necessary formulas and neglecting the penguin
contribution we are able to determine $A_0$, $A_2$, $\delta$ and
the value of the unitarity triangle angle $\alpha$ from the
experimental data on $B_{+-}$, $B_{00}$, $B_{+0}$ and $S_{+-}$,
which are presented in Table 1. By definition: $$B_{+-} \equiv
1/2[{\rm Br}(B_d \to\pi^+ \pi^-) + {\rm Br}(\bar B_d \to \pi^+
\pi^-)] \;\; ,$$ $$B_{00} \equiv 1/2[{\rm Br}(B_d \to\pi^0 \pi^0)
+ {\rm Br}(\bar B_d \to \pi^0 \pi^0)] \;\; ,$$ $$B_{+0}= {\rm
Br}(B_u \to\pi^+ \pi^0) = {\rm Br}(\bar B_u \to \pi^- \pi^0) \;\;
, $$ the last equality holds as far as the electroweak penguins
are neglected.

To extract the product $A_2 f_+(0)$ from $B_{+0}$ we will
use the value of $|V_{ub}|$ obtained from the general fit of the
Wolfenstein parameters of CKM matrix (CKM fitter, summer 2005): $
A = 0.825 \pm 0.019 \; , \;\; \lambda = 0.226 \pm 0.001 \; , \;\;
\bar\rho = 0.207 \pm 0.040 \; , \;\; \bar\eta = 0.340 \pm 0.023
\;\; : $
\begin{equation}
|V_{ub}| = (3.90 \pm 0.10) \cdot 10^{-3} \;\; . \label{4}
\end{equation}

From  (\ref{3}) and the experimental data on $B_{+0}$ from the
last column of Table 1 we readily get:
\begin{equation}
A_2 f_+(0) = 0.35 \pm 0.02 \;\; . \label{5}
\end{equation}

In order to understand if the factorization works in $B_u \to
\pi^+ \pi^0$ decay we should determine the value of $f_+(0)$. We
find it using the data on $B\to \pi l\nu$ decay from \cite{SL}:
\begin{equation}
f_+(0) = 0.22 \pm 0.02 \;\; , \label{6}
\end{equation}
thus getting:
\begin{equation}
A_2 = 1.60 \pm 0.20 \;\; , \label{7}
\end{equation}
which is not far from  the result of factorization:
\begin{equation}
A_2^f = \frac{8}{3\sqrt 3} (c_1 + c_2) \equiv \frac{2}{\sqrt
3}(a_1 + a_2) = 1.35 \;\; , \label{8}
\end{equation}
see Appendix. We come to the same conclusion as the authors of
paper \cite{Hod}: $A_2$ is estimated correctly by factorization.
Neglecting the penguin contribution we are able to extract the
values of $A_0$ and FSI phases difference $\delta$ from Eqs.
(\ref{1})-(\ref{3}) and the experimental data for $B_{+-}$,
$B_{00}$ and $B_{+0}$ from the last column of Table 1. In this way
we obtain:
\begin{equation}
A_0 = 1.53 \pm 0.23 \;\; , \label{9}
\end{equation}
which should be compared with the result of factorization:
\begin{equation}
A_0^f = \frac{\sqrt 2}{3\sqrt 3}(5 c_1 - c_2) = 1.54 \;\; ,
\label{10}
\end{equation}
see Appendix. In this way we come to the conclusion that
factorization works well for the moduli of both decay amplitudes.

For the phase difference $\delta \equiv \delta_2 - \delta_0$ we get:
\begin{equation}
\cos\delta = \frac{\sqrt 3}{4} \frac{B_{+-} -2B_{00} + \frac{2}{3}
\frac{\tau_0}{\tau_+} B_{+0}}{\sqrt{\frac{\tau_0}{\tau_+} B_{+0}}
\sqrt{B_{+-} + B_{00} - \frac{2}{3} \frac{\tau_0}{\tau_+} B_{+0}}}
\;\; , \label{11}
\end{equation}
\begin{equation}
\delta = \pm (53^o \pm 7^o) \;\; , \label{12}
\end{equation}
where $\tau_0/\tau_+ \equiv \tau(B_d)/\tau(B_u) = 0.92$ is
substituted. This is the place where the factorization which
predicts the negligible FSI phases fails.

In Section 4 we will present a model in which the pattern of
$B\to\pi\pi$ amplitudes obtained above is realized.

Let us turn to the bottom part of Table 1. Since we neglect
penguins the experimental value of $S_{+-}$ is directly related to
the unitarity triangle angle $\alpha$:
\begin{equation}
\sin 2\alpha^{\rm T} = S_{+-} \;\; , \label{13}
\end{equation}
$$\alpha_{\rm BABAR}^{\rm T} = 99^o \pm 5^o \; , \;\; \alpha_{\rm
Belle}^{\rm T} = 111^o \pm 6^o \; , \;\; \alpha_{\rm average}^{\rm
T} = 105^o \pm 4^o \;\; , $$ where index ``T'' stands for ``tree''
stressing that penguins are neglected (three other values of
$\alpha$ are not compatible with the Standard Model).

\section{FSI phases in $K\to \pi\pi$, $D\to\pi\pi$ and $B\to
D\pi$} The $s$-wave amplitudes of two pions production with $I=0$
and $I=2$ are generally different. In particular there are
quark-antiquark resonances in $s$-channel with $I=0$ but not with
$I=2$. This can lead to large difference of phases in the
channels with $I=0$ and $I=2$. Let us remind what experimental
data tell us about these phases at the pion center of mass
energies $E=m_K$ and $E=m_D$. Since at $E = m_K$ only elastic
rescattering of pions is possible (the inelastic channels are
closed since the energy is low) Watson theorem is applicable and
strong interaction phases of matrix elements of $K\to (2\pi)_I$
decays are equal to the  phases of amplitudes describing $\pi\pi
\to \pi\pi$ scattering at $E = m_K$. From the analysis  of $\pi\pi
\to \pi\pi$ scattering data performed in \cite{Choe} at
$E_{\pi\pi} = m_K$ we have:
\begin{equation}
\delta_0^K = 35^o \pm 3^o \;\; , \label{14}
\end{equation}
\begin{equation}
\delta_2^K = -7^o \pm 0.2^o \;\; , \label{15}
\end{equation}
\begin{equation}
\delta_0^K - \delta_2^K = 42^o \pm 4^o \;\; . \label{16}
\end{equation}
The large value of $\delta_0^K$ is due to the specific behaviour of
$I=0$, $J=0$ $\pi\pi$-phase attributed to the $f_0(600)$ (or
$\sigma$) ``resonance''.

The same value of the difference $\delta_0^K - \delta_2^K$ follows
from the analysis of $K_S \to \pi^+ \pi^-$, $K_S \to\pi^0\pi^0$
and $K^+ \to \pi^+\pi^0$ decay probabilities analogous to one we
perform for $B \to\pi\pi$ decays in Section 2 neglecting the penguin
contributions. (In case of $K \to \pi\pi$ decays the penguins are very
important being responsible for the enhancement of $I=0$
amplitude. Since CKM phase of the penguin amplitudes is almost the same
as that of the tree amplitude, the analysis performed
in Section 2 is applicable for kaon decays but the amplitude
$A_0^K$ should contain the penguin contribution as well.)

What concerns the moduli of the kaon decay amplitudes with $I=0$
and $I=2$, they are given with rather good accuracy (within 50\%
from the experimental data) by factorization \cite{SVZV}.

In case of $D\to\pi\pi$ decays, the gluon penguin amplitudes are
negligible in comparison with the tree ones (since the loop with
$s$-quark is subtracted from the one with $d$-quark while the momentum
transfer is of the order of $m_D^2$), and the effective
Hamiltonian responsible for these decays looks like:
\begin{eqnarray}
\hat{H}_D & = & \frac{G_F}{\sqrt 2} \sin\theta_c [c_1^D \bar u
\gamma_\alpha(1+\gamma_5)d \bar d \gamma_\alpha(1+\gamma_5)c +
\nonumber \\ & + & c_2^D \bar d \gamma_\alpha(1+\gamma_5)d \bar u
\gamma_\alpha(1+\gamma_5)c] \;\; , \label{17}
\end{eqnarray}
where $\theta_c$ is Cabibbo angle, $\sin\theta_c = 0.22$, $c_1^D
\approx 1.22$, $c_2^D \approx -0.42$ \cite{Bur}.

Calculating the matrix elements in the factorization approximation
we obtain:
\begin{equation}
M_{D\to\pi^+\pi^-} = \frac{G_F}{\sqrt 2} \sin\theta_c f_+^D(0) f_\pi
m_D^2 (c_1^D + \frac{c_2^D}{3}) \;\; , \label{18}
\end{equation}
\begin{equation}
M_{D\to\pi^0\pi^0} = \frac{G_F}{\sqrt 2} \sin\theta_c f_+^D(0)
f_\pi m_D^2 (\frac{c_1^D}{3}+ c_2^D) \;\; , \label{19}
\end{equation}
\begin{equation}
M_{D^\pm\to\pi^\pm\pi^0} = \frac{G_F}{\sqrt 2} \sin\theta_c
f_+^D(0) \frac{f_\pi}{\sqrt 2} m_D^2 \frac{4}{3}(c_1^D +
c_2^D) \;\; , \label{20}
\end{equation}
(the analogous formulas for $B$ decays are derived in Appendix)
while the isotopic analysis gives:
\begin{equation}
M_{D\to\pi^+\pi^-} = \frac{G_F}{\sqrt 2} \sin\theta_c f_+^D(0) f_\pi
m_D^2 (\frac{e^{i\delta_d}}{2\sqrt 3} A_2^D + \frac{1}{\sqrt 6}
A_0^D) \;\; , \label{21}
\end{equation}
\begin{equation}
M_{D\to\pi^0\pi^0} = \frac{G_F}{\sqrt 2} \sin\theta_c f_+^D(0)
f_\pi m_D^2 (\frac{e^{i\delta_d}}{\sqrt 3} A_2^D - \frac{1}{\sqrt
6} A_0^D) \;\; , \label{22}
\end{equation}
\begin{equation}
M_{D^\pm\to\pi^\pm\pi^0} = \frac{G_F}{\sqrt 2} \sin\theta_c f_+^D(0)
f_\pi m_D^2 (\frac{e^{i\delta_d}\sqrt 3}{2\sqrt 2} A_2^D) \;\; .
\label{23}
\end{equation}

From the recent study of semileptonic $D$-meson decays $D \to\pi
l\nu$ it was found \cite{Wi}:
\begin{equation}
f_+^D(0) = 0.62 \pm 0.04 \;\; . \label{24}
\end{equation}
Comparing (\ref{23}), (\ref{24}) and recent measurement \cite{Cl}:
\begin{equation}
{\rm Br}(D^\pm \to \pi^\pm \pi^0) = (1.25 \pm 0.10) \cdot 10^{-3}
\label{25}
\end{equation}
we obtain: $$ A_2^D = 0.88 \pm 0.08 \;\; , $$ which is not so
different from the factorization result,  (\ref{20}):
\begin{equation}
(A_2^D)_f = \frac{8}{3\sqrt 3} (c_1^D + c_2^D) = 1.2 \;\; .
\label{26}
\end{equation}

Comparing Eqs. (\ref{18})-(\ref{19}) with Eqs.
(\ref{21})-(\ref{22}) we obtain in the factorization
approximation:
\begin{equation}
(A_0^D)_f = \frac{\sqrt 2}{3\sqrt 3}(5c_1^D - c_2^D) = 1.8 \;\; ,
\label{27}
\end{equation}
while according to \cite{Cl} from the experimental data it follows:
\begin{equation}
A_0^D = 2.1 \pm 0.2 \;\; . \label{29}
\end{equation}
We see that the factorization results are within 30\% from the
experimental values of the moduli of the decay amplitudes.
However, factorization fails completely in describing the
difference of FSI phases. The data on $D\to\pi^+ \pi^-$, $D \to
\pi^0 \pi^0$ and $D^\pm \to \pi^\pm \pi^0$ branching ratios lead
to \cite{Cl}:
\begin{equation}
\delta_2^D - \delta_0^D \equiv \delta_D = \pm(86^o \pm 4^o) \;\; ,
\label{30}
\end{equation}
which is responsible for (or follows from) the relatively large $D\to
\pi^0 \pi^0$ decay probability  \cite{Cl}:
\begin{equation}
{\rm Br}(D^0 \to \pi^0 \pi^0)_{\rm exp} = (0.79 \pm 0.08) \cdot
10^{-3} \;\; , \label{31}
\end{equation}
of the order of that into $\pi^+ \pi^-$ \cite{Ei}: $$ {\rm Br}
(D^0 \to\pi^+ \pi^-)_{\rm exp} = (1.39 \pm 0.07) \cdot 10^{-3}
\;\; . $$ Using $\tau_{D_0}/\tau_{D_+} = 410/1040$ from \cite{Ei}
we readily reproduce the phase difference given by  (\ref{30})
with the help of  (\ref{11}).

In factorization approximation neglecting $\delta_D$ we will get:
\begin{eqnarray}
{\rm Br}(D^0 \to \pi^0 \pi^0)_f = \frac{1}{2}
\left[\frac{\frac{1}{3} c_1^D + c_2^D}{\frac{2\sqrt 2}{3} (c_1^D +
c_2^D)} \right]^2 \times \nonumber\\ \times {\rm Br}(D^+ \to
\pi^\pm \pi^0) \frac{\tau_{D_0}}{\tau_{D_+}} << {\rm Br}(D^0 \to
\pi^0 \pi^0)_{\rm exp} \;\; . \label{32}
\end{eqnarray}
The analogous
phenomena we encountered in $B\to\pi\pi$ decays.

Let us note that the $s$-wave resonance with zero isospin
 $f_0(1710)$ alone
cannot explain such a big phase; its contribution to $D \to
(\pi\pi)_{I=0}$ decay amplitude is proportional to:
\begin{equation}
m_D - m_{f_0} - i \frac{\Gamma_{f_0}}{2} = 150 \; {\rm MeV} - i 70
\; {\rm MeV} \; , \;\; |\delta_D| \approx 30^o \;\; .  \label{33}
\end{equation}

It is not easy to reconcile reasonable (20\%$\div$30\%) accuracy
of factorization in describing the moduli of the decay amplitudes
into $\pi\pi$ states with a definite isospin and the large FSI
phases difference since the latter signal of strong rescattering
of pions at $E \approx m_D$ which should not only generate phases
but also shift the moduli of the amplitudes. The resolution may be
that the interactions are ``semistrong'' in both channels: one
half of 86$^o$ comes from $I=0$, another from $I=2$ (just as in
the case of $B\to\pi\pi$ decays, see Section 4).

If we suppose that FSI phases scale with decaying meson mass as
$1/M$ we will get about $30^0$ phases difference for $B \to \pi
\pi$ decays from  (29).

$\pi\pi$ FSI phase shifts at $E=m_K$ and $E = m_D$ are not small.
However, in both cases we are in the regions where two pion
resonances are situated, which is not the case for the high energy
of the order of $B$-meson mass.

 Our last example is $B\to D\pi$ decays, where the energy is
high and we are definitely above the resonances domain, though the
FSI phase shift is nevertheless large \footnote{We are grateful to
A.E. Bondar who brought this case to our attention.}. $D\pi$ pair
produced in $B$-decays can have $I=1/2$ or $3/2$. From the
measurement of the probabilities of $B^- \to D^0 \pi^-$, $B^0 \to
D^- \pi^+$ and $B^0 \to D^0 \pi^0$ decays in paper \cite{Cleo} the
FSI phases difference of these two amplitudes was determined:
\begin{equation}
\delta_{D\pi} = 30^o \pm 7^o \;\; . \label{34}
\end{equation}

Concluding this section we wish to note that the  direct CP
asymmetry observed in $B_d(\bar B_d) \to \pi^{\mp} K^{\pm}$
decays is incompatible with small FSI phase difference between
$I=1/2$ and $I=3/2$ amplitudes.

\section{FSI phases: theoretical considerations}

There are many theoretical papers on
the final state interaction (FSI) in the
heavy-meson decays \cite{ka} - \cite{Al}. For example in paper \cite{CCS}
the final state
interactions in $B \to \pi \pi$ decay are modelled as the soft
rescattering of the certain intermediate two-body hadronic
channels ($\pi \pi, \rho
\rho, D^{*}\bar{D^{*}}, D\bar{D}$).
The hadronic amplitudes, which enter the calculation of the
imaginary parts of the decay amplitudes
were described by   $\pi, \rho, D, D^{*}
$-meson exchanges in the $t$-channel. Rather large phases due to FSI have
been
obtained. While for $\pi$-exchange (with the pole close to the physical region)
this procedure is reasonable, it exaggerates
the
contributions of the vector exchanges ($\rho, D^{*}$), which for the
elementary
particle exchange with spin $J=1$ gives the partial wave amplitude,
which does not
decrease with energy.
In reality all the exchanges should be reggeized and in the
physical region of the processes corresponding intercepts $\alpha_i (0) < 1$
(for
$D^{*}$ the most probably value of the intercept is negative). This will lead
to
the strong reduction of the corresponding amplitudes (see for example
\cite{De}).

A number of papers (\cite{De} - \cite{Al}) use Watson theorem in order
to extract the
 phases of the
decay amplitudes by multiplying the bare matrix elements by
$S^{1/2}$ (where $S$ is the $S$-matrix). However in $B$-decays
there are many coupled multiparticle channels.  In this case such
a procedure can be applied only in the basis of the eigenstates
which diagonalize $S$-matrix. But for the realistic strong
interactions this is impossible at present.

Another approximation is to use the Feynman diagrams approach
taking only the low mass intermediate states $X, Y$ into account.
This approach coincides with the use of the unitarity condition
only if the transitions $\pi \pi \to X Y$ are described by the
real amplitudes. This is certainly not true for elastic $\pi
\pi$-scattering, where the amplitude at  large energies is
predominantly imaginary. In this formalism the resulting decay
matrix elements are:

\begin{equation}
M^{I}_{\pi \pi} = M^{(0) I}_{XY} \left(\delta_{\pi X}
\delta_{\pi Y}  + i
T^{J=0}_{XY \to \pi \pi} \right) \;\;  , \label{109}
\end{equation}
where  $M^{(0) I}_{XY}$ are the decay matrix elements without FSI
and $T^{J=0}_{XY \to \pi \pi}$ is the $J=0$ partial wave amplitude
of the process $XY \rightarrow \pi \pi$\footnote{We use the
standard normalization with $T^J = \frac{S^J - 1}{2 i}$.}. At very
high energies the amplitudes of $\pi \pi$ elastic scattering are
imaginary and  $T^{J=0}_{XY \to \pi \pi}$     do not decrease with
energy (mass of a heavy meson). Thus this contribution, according to
(\ref{109}) does not change
 the phase of the matrix element, but only changes its modulus.
 The extra phases come from the real parts of the
amplitudes, which in Regge model
are
due to the secondary exchanges $(R \equiv \rho, f, ...)$, which decrease with
 energy as $1/s^{1-\alpha_{R} (0)} \approx 1/\sqrt{s}$.
The contribution of the pion exchange in the $t$-channel, which
 is dominant in the process $\rho\rho\to \pi\pi$\footnote{ In
 $B$-decays transverse polarizations of $\rho$-mesons are small
 that is why $a_2$ and $\omega$ exchanges in $\rho\rho\to \pi\pi$
 amplitudes are suppressed.} decreases even faster (as $1/s$).
 However ${\rm Br}(B\to \rho^+\rho^-)$ is substantially larger than ${\rm Br} (B\to\pi^+\pi^-)$ and $\rho\rho$  intermediate state is important
 in  (\ref{109}). This is especially true for color suppressed $B\to
 \pi^o\pi^o$ decay, where the chain $B\to \rho^+\rho^- \to
 \pi^o\pi^o$ is enhanced. On the contrary $\pi\pi$ contribution
 in $B\to \rho\rho$ decay is relatively suppressed. Using
 Regge analysis of $\pi\pi$  scattering \cite{BGK} and $\pi$-exchange
 model for $\rho\rho\to\pi\pi$ transitions, we obtain phases
  due to final state interactions for $\pi\pi$ final state
$\delta_2 \approx -12^0$
and $\delta_0 \approx 18^o $.
Thus the phase difference $\delta \approx
  -30^o$ is generated by intermediate $\rho \rho$
and $\pi \pi $ states
\footnote {An accuracy of this number is about $15^o$.}.

Sign of $\delta$ is negative, just as in the case of $K \to \pi
\pi$ decays. In this way in the numerical estimates we will use
negative value of $\delta$ from  (\ref{12}):
\begin{equation}
\delta \approx -(50 \pm 7)^o \;\; , \; \delta_0 = 30^o \;\; ,
 \; \delta_2 = -20^o \;\; .
\label{355}
\end{equation}

As far as ${\rm Br}(B\to \rho^+\rho^-)$ is much larger than ${\rm
Br}
 (B\to\pi^+\pi^-)$ because of enhancement in tree amplitudes but
not in penguin amplitudes (contribution of penguins in
 $B\to \rho\rho$ amplitude is small) we should expect
$\tilde\delta_0$ to be substantially smaller than $\delta_0$.

Note that in this model there is little change in
  moduli of amplitudes in comparison with factorization
  predictions.

For $B\to \rho\rho$ decays the same model gives
  $\approx -5^o$ for $I=2$ and $+5^o$ for $I=0$ amplitudes,
  resulting in a small phase difference $\approx 10^o$, consistent
  with experiment.

 Thus the lowest mass hadronic intermediate states may produce the phases
which are consistent with the data on $B \to \pi \pi$ and $B \to
\rho \rho$ decays. There are many high-mass states as well, and
they can lead to additional phases (the inclusion of $\pi a_1$
intermediate state makes $\delta \approx -40^o$).

\section{Taking penguins into account: shifts of $A_0$, $\delta$
and $\alpha$ and the values of $C_{+-}$ and $C_{00}$}

Let us analyse to what changes of the parameters introduced and
calculated in Section 2 penguins lead. Since QCD penguins
contribute only to $I=0$ amplitude the value of $A_2$ extracted
from $B_{+0}$ remains the same, see (\ref{7}). The requirement
that the numerical values of $B_{+-}$ and $B_{00}$ are not shifted
when penguins are taken into account leads to the following shifts
of the amplitude $A_0$ and phase difference $\delta$:
\begin{equation}
A_0 \to A_0 + \tilde{A}_0 \; , \;\; \delta \to \delta +
\tilde{\delta} \;\; ,  \label{35}
\end{equation}
\begin{equation}
\tilde A_0 = \sqrt 6 \left|\frac{V_{td}}{V_{ub}}\right| \cos\alpha
\cos(\delta_p + \tilde\delta_0 - \delta_0) P \;\; , \label{36}
\end{equation}
\begin{equation}
\tilde\delta = -\sqrt 6 \left|\frac{V_{td}}{V_{ub}}\right|
\cos\alpha \sin( \delta_p + \tilde\delta_0 - \delta_0)  P/A_0 \;\; , \label{37}
\end{equation}
where only the terms linear in $P$ are taken into account. For
numerical estimates we take:
\begin{equation}\left|\frac{V_{td}}{V_{ub}}\right| =
\frac{\sin\gamma}{\sin\beta} \approx 2.3 \pm 0.2 \;\; , \label{38}
\end{equation}
where $\beta = 22^o$, $\gamma = 60^o \pm 10^o$. In the
factorization approach we have (see Appendix)\footnote {Note that
the definition of $P$ used in the present paper differs in sign
from that in \cite{V}.} :
\begin{equation}
P^f = -a_4 - \frac{2m_\pi^2}{(m_u + m_d) m_b} a_6 = 0.06 \;\; ,
\label{39}
\end{equation}
and shifts of $A_0$ and $\delta$ are small:
\begin{equation}
-0.12 < \tilde A_0 < 0.12 \; , \;\; -4^o < \tilde\delta < 4^o
\label{40}
\end{equation}
for
\begin{equation}
A_0 = 1.5 \; , \;\; -1 < \cos( \delta_p + \tilde\delta_0 - \delta_0) \; , \;\;
 \sin ( \delta_p + \tilde\delta_0 - \delta_0)< 1
\;\; {\rm and} \;\; 70^o < \alpha < 110^o \;\; . \label{41}
\end{equation}
In particular even if the penguin contribution is underestimated
by factor 2, the statement that $\delta + \tilde\delta$ is large
still holds\footnote{Indication of such an underestimate follows
from the probability of $b\to s$ penguin dominated $B^+ \to K^0
\pi^+$ decay.} (note that $\alpha$ can be closer to 90$^o$).

The following two equations for direct CP asymmetries
determine $P$ and $ \delta_p + \tilde\delta_0 - \delta_0$
(as far as $A_0, \; A_2$ and $\delta = \delta_2 - \delta_0$ are
known):
\begin{eqnarray}
C_{+-} & = & -\frac{\tilde P}{\sqrt 3} \sin\alpha [\sqrt 2 A_0
\sin(\delta_0 - \tilde\delta_0 - \delta_p) + A_2 \sin(\delta_2 -
\tilde\delta_0 -\delta_p)]/ \nonumber \\ & / &[\frac{A_0^2}{6} +
\frac{A_2^2}{12} + \frac{A_0 A_2}{3\sqrt 2} \cos\delta -
\sqrt{\frac{2}{3}} A_0 \tilde P \cos\alpha \cos(\delta_0 -
\tilde\delta_0 - \delta_P) - \nonumber \\ & - & \frac{A_2 \tilde
P}{\sqrt 3} \cos\alpha \cos(\delta_2 - \tilde\delta_0 - \delta_p)
+ \tilde P^2] \;\; , \label{42}
\end{eqnarray}
\begin{eqnarray}
C_{00} & = & -\sqrt{\frac{2}{3}} \tilde P \sin\alpha [A_0
\sin(\delta_0 - \tilde\delta_0 -\delta_p) - \sqrt 2 A_2
\sin(\delta_2 - \tilde\delta_0 -\delta_p)] / \nonumber \\
& / & [\frac{A_0^2}{6} + \frac{A_2^2}{3} - \frac{\sqrt 2}{3} A_0 A_2
\cos\delta - \sqrt{\frac{2}{3}} A_0 \tilde P \cos\alpha
\cos(\delta_0 - \tilde\delta_0 - \delta_p) + \nonumber \\ & + &
\frac{2}{\sqrt 3} A_2 \tilde P \cos\alpha \cos(\delta_2 -
\tilde\delta_0 -\delta_p) + \tilde P^2] \;\; , \label{43}
\end{eqnarray}
where
\begin{equation}
\tilde P \equiv \left|\frac{V_{td}^* V_{tb}}{V_{ub}
V_{ud}^*}\right| P \approx 2.3 P \;\; . \label{44}
\end{equation}

Three last terms in denominators of (\ref{42}) and (\ref{43}) lead
to less than 10\% variations of the numerical values of $C_{+-}$
and $C_{00}$ for $\tilde P < 0.3$. Neglecting them we get:
\begin{equation}
\frac{\sin(\delta_2 - \tilde\delta_0 -\delta_p)}{\sin(\delta_0 -
\tilde\delta_0 -\delta_p)} = \frac{1-0.57 \frac{C_{00}}{C_{+-}}}
{1.4+0.41 \frac{C_{00}}{C_{+-}}} \;\; , \label{45}
\end{equation}
where the numerical values for $A_2$, $A_0$ and $\delta$ from
(\ref{7}), (\ref{9}) and (\ref{12}) correspondingly were used.

From the central values in the last column in Table 1
of $C_{+-}$ and $C_{00}$ we get:
\begin{equation}
\delta_0 -\tilde\delta_0 - \delta_p = 70^o \;\; , \;\; P = 0.11
\;\; . \label{46}
\end{equation}

The numerical value of $P$ is two times larger than the
factorization estimate of it presented in (\ref{39}), while
$\delta_0 -\tilde\delta_0 - \delta_p$ largely deviates from $30^o$
which is our estimate of $\delta_0$, while  $\tilde\delta_0$
should be considerably smaller as well as $\delta_p$ the latter
being close to $30^o$ only for very asymmetric configurations of
quarks in $\pi$ mesons and is smaller otherwise \cite{BSS}. If the
experimental accuracy of $C_{ik}$ were  good we would be able to
use the results obtained for determination of the value of the
angle $\alpha$ from $S_{+-}$, realizing in this way Gronau-London
approach \cite{GL}.

However the experimental uncertainty in $C_{00}$ is very
big, while the measurements of $C_{+-}$ by Belle and BABAR contradict
each other. So let us look which values of the direct asymmetries
follow from our formulas.

Denominator of the expression for $C_{+-}$ is close to one and in
the expression in brackets in nominator first term dominates.
Neglecting $\tilde\delta_0$ and $\delta_p$ and taking $\delta_0 =
30^o$, $\delta_2 = -20^o$ we get $C_{+-}=-0.04$ for the value of
penguin amplitude obtained in the factorization approach,
(\ref{39}). We reproduce BABAR central value of $C_{+-}$ if we
suppose that factorization underestimate penguin amplitude by
factor 2; however in order to reproduce Belle number we should
accept that factorization is wrong by factor 10, which looks
highly improbable.

What to do if $C_{+-}$ appeared to be equal to the average of the
present day Belle and BABAR results $C_{+-} \approx -0.3$? One
possibility is to suppose that $\delta_2 - \tilde\delta_0
-\delta_P \approx 0$, while $\delta_0 - \tilde\delta_0 - \delta_P
\approx 50^o$ and to look for FSI mechanism which provides such a
result\footnote{Let us note that in paper \cite{GR} argument in 
favor of $C_{+-} \approx -0.3$ is presented which is based on the
comparison of the direct CP violation in $B(\bar B) \to \pi^+\pi^-$
and $B(\bar B) \to K^+\pi^-(K^-\pi^+)$ decays (see also \cite{Haz}).}.

$C_{00}$ is also negative while its absolute value is larger
than
 $C_{+-}$: denominator is about .55 while in nominator both terms are
negative.



The requirement that the value of CP asymmetry $S_{+-}$ is not changed
when penguins are taken into account leads to the following shift
of the value of the unitarity triangle angle $\alpha$:
\begin{equation}
\alpha = \alpha^T + \tilde\alpha \;\; , \label{50}
\end{equation}
\begin{equation}
\tilde\alpha =  -\frac{\tilde P}{2 \sqrt 3} \sin\alpha [\sqrt 2 A_0
\cos(\delta_0 - \tilde\delta_0 - \delta_P) + A_2 \cos(\delta_2 -
\tilde\delta_0 -\delta_P)] \;\;
. \label{51}
\end{equation}
Substituting the numerical values of $A_2$ from (\ref{7}), $A_0$
from (\ref{9}), $\tilde P$ from (\ref{44}) and substituting
$\sin\alpha$ by one and both cos by $0.9$ we get:
\begin{equation}
\tilde\alpha \approx -2.3 P \approx -7^o, \label{53}
\end{equation}
where the result of the matrix element of the penguin operator
calculation in factorization approximation (\ref{39}) is used. We
observe that our approach is at least selfconsistent: the shift of
$\alpha$ due to penguin contribution is small. For the BABAR
 value of $S_{+-}$ we obtain:
\begin{equation}
\alpha_{\rm BABAR}
= \alpha_{\rm BABAR}^T + \tilde\alpha = 92^o \pm 5^o
\;\; . \label{54}
\end{equation}
In the case of the averaged experimental values we get:
\begin{equation}
\alpha_{\rm average} =
\alpha_{\rm average}^T + \tilde\alpha = 98^o \pm
4^o \;\; . \label{55}
\end{equation}

Theoretical uncertainty of the value of $\alpha$ can be estimated
in the following way. Let us suppose that the accuracy of the
factorization calculation of the penguin amplitude is 100\% (in
all the examples considered in this paper it was much better).
Then:
\begin{equation}
\tilde\alpha = -7^o \pm 7^o_{\rm theor} \;\; ,
\label{56}
\end{equation}
\begin{equation}
\alpha_{\rm average} = 98^o \pm 4^o_{\rm exp} \pm 7^o_{\rm theor}
\;\; , \label{57}
\end{equation}
while BABAR value is smaller:
\begin{equation}
\alpha_{\rm BABAR} = 92^o \pm 5^o_{\rm exp} \pm 7^o_{\rm theor}  \;\; .
\label{58}
\end{equation}

Better theoretical accuracy of $\alpha$ follows from $B \to \rho^+
\rho^-$ decays, where penguin contribution is two times smaller.
Since FSI phases are small in these decays, results of the paper
\cite{V}  are directly applicable:
\begin{equation}
\alpha^{\rho \rho} = 92^o \pm 7^o_{\rm exp} \pm 4^o_{\rm
theor} \;\; ,
\label{111}
\end{equation}
where we take the WHOLE penguine contribution as an estimate of the
theoretical uncertainty.

The model independent isospin analysis of $ B \to \rho \rho $ decays performed
by BABAR gives \cite{BA} :
\begin{equation}
\alpha^{\rho \rho}_{BABAR} = 100^o \pm 13^o \;\; ,
\label{59}
\end{equation}
while the analogous analysis performed by Belle gives \cite{Be} :
\begin{equation}
\alpha^{\rho \rho}_{Belle} = 87^o \pm 17^o \;\; .
\label{60}
\end{equation}

Finally, the global CKM fit results are\cite{CKM,UT}:

\begin{equation}
\alpha_{CKMfitter} = 97^o \pm 5^o \;\;  , \;\;  \alpha_{UTfit} =
95^o \pm 5^o\;\; . \label{61}
\end{equation}

\section{Conclusions}

\begin{enumerate}
\item The moduli of the amplitudes $A_0$ and $A_2$ of B decays into
$\pi \pi$ states are given with good accuracy by factorization of
the tree quark diagram, while FSI phase shift between these two
amplitudes is very large, $|\delta| \approx 50^o$, which explains
large $B_d \to\pi^0\pi^0$ decay probability.
\item Theoretical uncertainty of the value of $\alpha$ extracted
from $B \to\pi\pi$ data on $S_{+-}$ is at the level of few
degrees.
\item Resolution of the contradiction of Belle and BABAR
experimental data on CP asymmetries in $B\to\pi\pi$ decays is very
important both for understanding the FSI dynamics (the data on
$C$) and for determination of angle $\alpha$ (the data on $S$).
\end{enumerate}

We are grateful to A.E. Bondar, A.Yu. Khodjamirian and L.B. Okun
for useful discussions.

 A.K. was partly supported by grants
CRDF RUP 2-2621-M0-04 and RFBR 04-02-17263;
M.V. was partly supported by grants RFBR 05-02-17203 and
NSh-5603.2006.2.

\newpage

\begin{center}

{\bf Table 1.} Experimental data on $B\to\pi\pi$ decays. Branching
ratios are in units of $10^{-6}$.

\bigskip

\begin{tabular}{|l|c|c|c|} \hline
& BABAR & Belle & Heavy Flavor \\ & & & Averaging Group \cite{hf}
\\ \hline $B_{+-}$ & $5.5 \pm 0.5$ & $4.4 \pm 0.7$ & $5.0 \pm 0.4$
\\ $B_{00}$ & $1.17 \pm 0.33$ & $2.3 \pm 0.5$ & $1.45 \pm 0.29$ \\
$B_{+0}$ & $5.8 \pm 0.7$ & $5.0 \pm 1.3$ & $5.5 \pm 0.6$ \\ \hline
$S_{+-}$ & $-0.30 \pm 0.17$ & $-0.67 \pm 0.16$ & $-0.50 \pm 0.12$
\\ $C_{+-}$ & $-0.09 \pm 0.15$ & $-0.56 \pm 0.13$ & $-0.37 \pm
0.10$ \\ $C_{00}$ & $-0.12 \pm 0.56$ &$-0.44 \pm 0.56$ & $-0.28
\pm 0.39$ \\ \hline
\end{tabular}
\end{center}

\newpage

\begin{center}

{\bf Appendix}

\end{center}

\setcounter{equation}{0} \def\theequation{A\arabic{equation}}

\bigskip

In order to calculate amplitude $A_2$ in the factorization
approximation we start from the following Hamiltonian:
\begin{eqnarray}
\hat H & = & \frac{G_F}{\sqrt 2} \left| V_{ub} V_{ud}^*\right|
e^{-i\gamma} [c_1 \bar u \gamma_\alpha (1+\gamma_5)b \bar d
\gamma_\alpha(1+\gamma_5)u + \nonumber \\ & + & c_2 \bar d
\gamma_\alpha (1+\gamma_5)b \bar u \gamma_\alpha (1+ \gamma_5)u]
\;\; , \label{A1}
\end{eqnarray}
and take the matrix element of it between $\bar B_u$ and $\pi^- \pi^0$
states:
\begin{eqnarray}
M^f_{\bar B_u \to \pi^- \pi^0} & = & \frac{G_F}{\sqrt
2}\left|V_{ub} V_{ud}^*\right| e^{-i\gamma} \left[(c_1 +
\frac{c_2}{3}) <\pi^- |\bar d \gamma_\alpha (1+ \gamma_5)u|0>
\right. \times \nonumber \\ & \times & <\pi^0 | \bar u
\gamma_\alpha (1+\gamma_5)b| \bar B_u> +(\frac{1}{3} c_1 + c_2)
<\pi^0 | \bar u \gamma_\alpha(1+\gamma_5)u|0> \times \nonumber \\
& \times & \left. <\pi^- | \bar d \gamma_\alpha (1+\gamma_5)b|
\bar B_u> \right] = \\ & = & \frac{G_F}{\sqrt 2} \left| V_{ub}
V_{ud}^*\right| e^{-i\gamma} \frac{4}{3}(c_1 + c_2) < \pi^- |\bar
d \gamma_\alpha (1+\gamma_5)u|0> \times \nonumber \\ & \times &
<\pi^0|\bar u \gamma_\alpha(1+\gamma_5)b|\bar B_u > =
\frac{G_F}{\sqrt 2} \left|V_{ub} V_{ud}^*\right| e^{-i\gamma}
\frac{4}{3}(c_1 + c_2) f_\pi P_\alpha^{\pi^-} \times \nonumber \\ &
\times & \frac{1}{\sqrt 2} f_+(0) (P_\alpha^B + P_\alpha^{\pi^0}) =
\nonumber \\ & = & \frac{G_F}{\sqrt 2} \left|V_{ub} V_{ud}^*
\right| e^{-i\gamma} f_\pi m_B^2 \frac{1}{\sqrt 2} f_+(0)
\frac{4}{3}(c_1 + c_2) \;\; . \nonumber \label{A2}
\end{eqnarray}
Comparing the last expression with (\ref{3}) we get:
\begin{equation}
A_2^f = \frac{8}{3\sqrt 3} (c_1 + c_2) = \frac{2}{\sqrt 3} (a_1 +
a_2) = 1.35 \;\; , \label{A3}
\end{equation}
where $a_1 \equiv c_1 + c_2/3 = 1.02$, $a_2 \equiv c_2 + c_1/3 =
0.15$ and the values $c_1 = 1.09$, $c_2 = -0.21$ from  paper
\cite{Bur} were used.

To find the amplitude $A_0$ in the factorization approximation let
us take the matrix element of $\hat H$ between $\bar B_d$ and
$\pi^+ \pi^-$-states:
\begin{equation}
M^f_{\bar B_d \to \pi^+ \pi^-} = \frac{G_F}{\sqrt 2} \left| V_{ub}
V_{ud}^*\right| e^{-i\gamma} f_\pi m_B^2 f_+(0) (c_1 +
\frac{c_2}{3}) \;\; , \label{A4}
\end{equation}
and comparing it with (\ref{1}) we get:
\begin{equation}
\frac{1}{2\sqrt 3} A_2^f + \frac{1}{\sqrt 6} A_0^f = (c_1 +
\frac{c_2}{3}) \;\; . \label{A5}
\end{equation}

From (\ref{A3}) and (\ref{A5}) we obtain:
\begin{equation}
A_0^f = \frac{\sqrt 2}{3\sqrt 3} (5 c_1 - c_2) = 1.54 \;\; .
\label{A6}
\end{equation}
The amplitude of $\bar B_d \to \pi^0 \pi^0$ decay can be
calculated analogously, or it can be constructed from $A_0^f$ and
$A_2^f$ with the help of (\ref{2}):
\begin{eqnarray}
M^f_{\bar B_d \to \pi^0 \pi^0} & = & \frac{G_F}{\sqrt 2}
\left|V_{ub} V_{ud}^*\right| e^{-i\gamma} f_\pi m_B^2 f_+(0)
\left[-\frac{5c_1 - c_2}{9} + \frac{8c_1 + 8c_2}{9}\right] =
\nonumber \\ & = & \frac{G_F}{\sqrt 2} \left|V_{ub}
V_{ud}^*\right| e^{-i\gamma} f_\pi m_B^2 f_+(0) (c_2 +
\frac{c_1}{3}) \;\; , \label{A7}
\end{eqnarray}
a well known prediction of smallness of $\bar B_d \to\pi^0 \pi^0$
decay probability in the factorization approximation. However if
FSI phase shift between $A_0$ and $A_2$ is large, this
compensation disappears.

QCD penguins generate an additional term to the weak interaction
Hamiltonian, (\ref{A1}):
\begin{equation}
\Delta \hat H = \frac{G_F}{\sqrt 2}(-) V_{tb} V_{td}^* \left[c_3
O_3 + c_4 O_4 + c_5 O_5 + c_6 O_6\right] \;\; , \label{A8}
\end{equation}
$$
\begin{array}{ll}
O_3 = \bar d \gamma_\alpha(1+\gamma_5)b \left[\bar u
\gamma_\alpha(1+\gamma_5)u + \bar d \gamma_\alpha(1+\gamma_5)d
\right] \; , & c_3 = 0.013 \; ,
\\ O_4 = \bar d_a \gamma_\alpha(1+\gamma_5)b^c \left[\bar u_c
\gamma_\alpha(1+\gamma_5)u^a + \bar d_c
\gamma_\alpha(1+\gamma_5)d^a\right] \; , & c_4 = -0.032 \; ,
\\ O_5 = \bar d \gamma_\alpha(1+\gamma_5)b \left[\bar u
\gamma_\alpha(1-\gamma_5) u + \bar d \gamma_\alpha(1-
\gamma_5)d\right] \; , & c_5 = 0.009 \; ,
\\ O_6 = \bar d_a \gamma_\alpha(1+\gamma_5)b^c \left[\bar u_c
\gamma_\alpha(1-\gamma_5)u^a + \bar d_c
\gamma_\alpha(1-\gamma_5)d^a\right] \; , & c_6 = -0.037 \; .
\end{array} $$

In order to find the penguin contribution to $B\to\pi\pi$ decay
amplitudes let us calculate the matrix element of (\ref{A8})
between $\bar B_d$ and $\pi^+ \pi^-$ in the factorization
approximation. For this purpose it is convenient to use Fierz
transformations of $\gamma$-matrices rewriting $\Delta \hat H$ in
the following form:
\begin{eqnarray}
\Delta \hat H & = & -\frac{G_F}{\sqrt 2} V_{tb} V_{td}^* \left[a_4
\bar u \gamma_\alpha (1+ \gamma_5)b \bar d \gamma_\alpha (1+
\gamma_5) u - \right. \nonumber \\ & - & \left. 2a_6 \bar u (1+
\gamma_5) b \bar d (1- \gamma_5) u \right] \;\; , \label{A9}
\end{eqnarray}
where $a_4 = c_4 + 1/3 c_3 = -0.027$, $a_6 = c_6 + 1/3 c_5 =
-0.034$.

Calculating the matrix element:
\begin{equation}
<\pi^+ \pi^- |\Delta \hat H|\bar B_d > = -\frac{G_F}{\sqrt 2}
|V_{tb} V_{td}^* | e^{i\beta} f_+(0) f_\pi m_B^2 \left[a_4 +
\frac{2m_\pi^2 a_6}{(m_u + m_d)(m_b - m_u)} \right] \;\; ,
\label{A10}
\end{equation}
and comparing this expression with (\ref{1}) we get:
\begin{equation}
P^f = -a_4 - \frac{2m_\pi^2}{(m_u + m_d)m_b} a_6 = +0.06 \pm 0.01
\;\; , \label{A11}
\end{equation}
where $m_u + m_d = 9 \pm 3$ MeV, $m_b = 4.5$ GeV were substituted.

\end{document}